# DOTIFS: Spectrograph optical and opto-mechanical design

Haeun Chung*[a,b], A. N. Ramaprakash[c], Pravin Khodade[c], Deepa Modi[c], Chaitanya V. Rajarshi[c],
Sabyasachi Chattopadhyay[c], Pravin A. Chordia[c], Vishal Joshi[d], Sungwook E. Hong[e],
Amitesh Omar[f], Swara Ravindranath[g], Yong-Sun Park[a], and Changbom Park[b]

[a]Seoul National University, Seoul, Korea
[b]Korea Institute for Advanced Study, Seoul, Korea
[c]Inter-University Centre for Astronomy and Astrophysics, Pune, India
[d]Physical Research Laboratory, Ahmedabad, India
[e]Korea Astronomy and Space Science Institute, Daejeon, Korea
[f]Aryabhatta Research Institute of Observational Sciences, Nainital, India
[g]Space Telescope Science Institute, Baltimore, United States

## ABSTRACT

Devasthal Optical Telescope Integral Field Spectrograph (DOTIFS) is a new multi-Integral Field Unit (IFU) instrument, planned to be mounted on the 3.6m Devasthal optical telescope in Nainital, India. It has eight identical, fiber-fed spectrographs to disperse light coming from 16 IFUs. The spectrographs produce 2,304 spectra over a 370-740nm wavelength range simultaneously with a spectral resolution of R=1200-2400. It is composed of all-refractive, all-spherical optics designed to achieve on average 26.0% throughput from the telescope to the CCD with the help of high transmission spectrograph optics, volume phase holographic grating, and graded coated e2v 2K by 4K CCD. We present the optical and opto-mechanical design of the spectrograph as well as current development status. Optics and opto-mechanical components for the spectrographs are being fabricated.

## 1. INTRODUCTION

Devasthal Optical Telescope Integral Field Spectrograph (DOTIFS)[1] is a new optical multi-Integral Field Unit (multi-IFU) spectroscopic instrument being built by Inter-University Centre for Astronomy and Astrophysics (IUCAA). It is planned to be mounted on 3.6m Devasthal Optical Telescope (DOT)[2] at Devasthal peak, Uttarakhand, India. The telescope is manufactured by Advanced Mechanical and Optical Systems (AMOS) and managed by Aryabhatta Research Institute of Observational Sciences (ARIES). Korea Institute for Advanced Study (KIAS) and Seoul National University (SNU) participate in this project as international collaborators. DOTIFS is a unique instrument that has the multi-IFU capability with a novel deployment system. It is designed for 2-dimensional spatially resolved spectroscopy in the entire visible range with medium spectral resolution on 16 objects simultaneously. Overview of the instrument, as well as its science cases, are described in the previous paper[1].

The overall structure of the instrument is shown in Figure 1. The light coming from the telescope is directed to the Cassegrain side port by telescope tertiary mirror. It passes DOTIFS fore-optics, which slows down F ratio of the light and form 8' magnified focal plane. IFU deployment system allocates 16 IFUs on the focal plane, and the IFUs gather light from the observation targets on the sky. Each IFU has 8.7" x 7.4" field of view with fiber coupled 12x12 hexagonal aperture shape microlens array. The other end of fibers is connected to the entrance of 8 identical spectrographs, and spectrographs disperse light from 370 to 740nm wavelength range simultaneously in a single exposure. Spectrographs are mounted on the telescope Cassegrain port rather than a gravitationally invariant area to minimize fiber length and therefore maximize the throughput. They occupy the remaining space around the telescope direct port instrument, ARIES Devasthal Faint Object Spectrograph Camera (ADFOSC) [3].

*hchung@astro.snu.ac.kr; phone +82-10-7542-2737

In this paper, we present spectrograph optics and opto-mechanical design. We start with the overall design of the spectrograph and introduce the optical design with design principles. We report the result of tolerance and thermal analysis and briefly introduce spectrograph opto-mechanical structure. We also present expected optical performance the spectrograph. Finally, the current status of the spectrograph and the overall instrument is reported, followed by a summary of the paper.

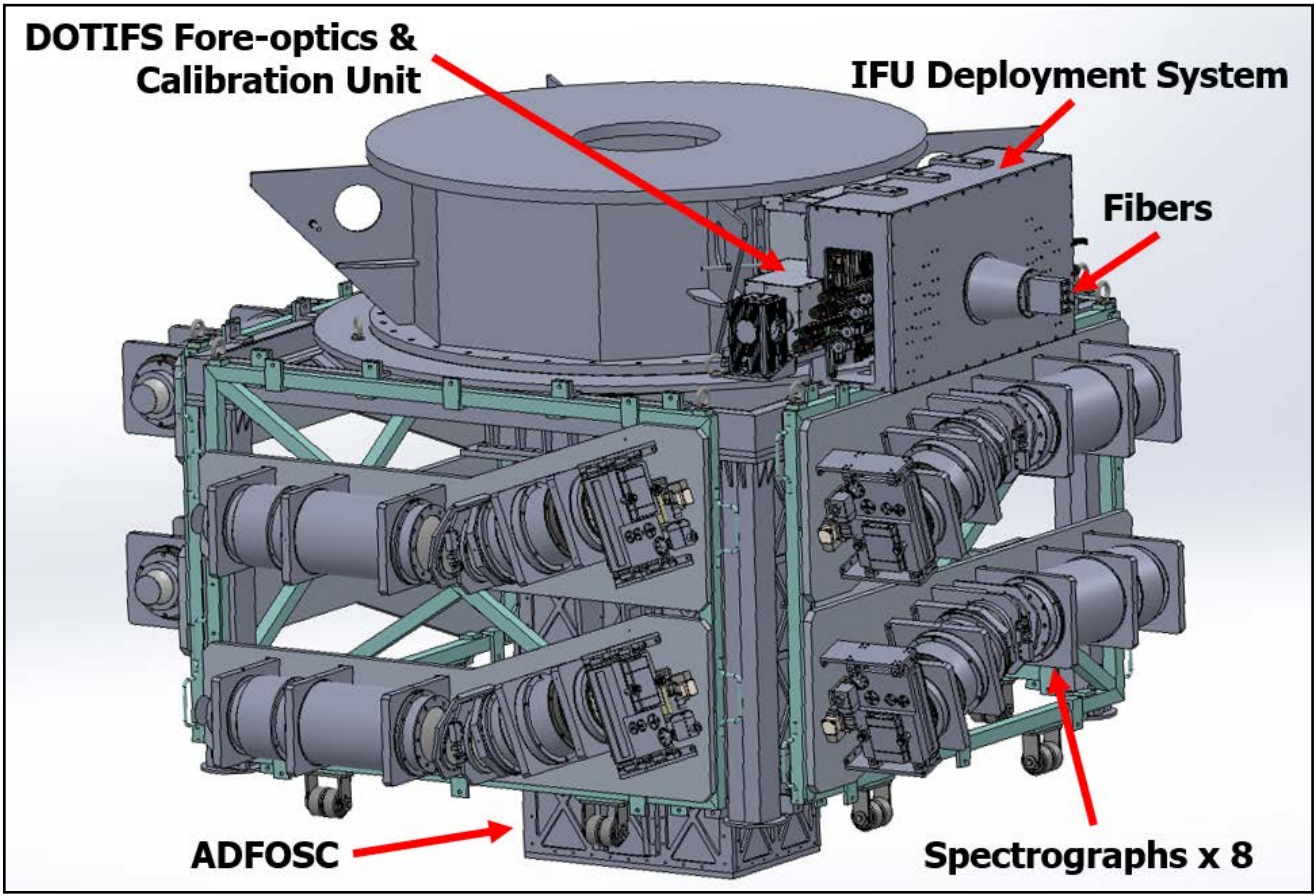

Figure 1 Overall mechanical rendering of DOTIFS mounted on the telescope. Each part of the spectrograph is pointed out with red arrows. 8 spectrographs are located around the main port instrument ADFOSC. 2 spectrographs are mounted at each side, in total four sides. Fibers between the IFU deployment system to each spectrograph are not shown.

## 2. OVERALL DESIGN OF THE SPECTROGRAPH

### 2.1 Design Goals
We describe design goals of the spectrograph optics below. We explain concerns or issues regarding each goal as well.

• Wavelength coverage from 370 nm to 740 nm

The wavelength coverage affects available optics material choices while designing optics. This range is mostly in the visible region, and there are diverse kinds of glass with good transmission in the visible range. However, there are only a few glasses with high transmission (> 99.5 %/10 mm) at near-UV range (370 to 400nm). A small difference in transmission (%/10 mm) ends up with a huge amount of light loss because the light loss is increasing exponentially as optics size becomes big. It leads us to use Calcium Fluoride ($CaF_2$) and other high UV throughput i-line glass materials which have transmission higher than 99.5 %/10 mm. An option for splitting wavelength range with a dichroic element was rejected since a single-arm spectrograph can satisfy the required wavelength range and spectral resolution.

- Spectral resolution of R~1500-1800

The requirement of R~1800 at the central wavelength demands a dispersion of 0.00822 nm/μm, assuming 2.5 pixels sampling per spectral element. 2.5 pixels corresponds to 37.5 μm with 15 μm size pixel, and this will be the size of fiber image on the detector. Since the diameter of the fiber is specified as 100 μm and an angle of light coming out of fiber is about F/4, it leads spectrograph optics to have F/4 collimator and F/1.5 camera. Wavelength coverage from 370 nm to 740 nm corresponds to 3000 pixels on the detector with required dispersion.

- > 80% Ensquared energy within 2.5 pixels from monochromic 100 μm size, circular, and NA=0.125 light.

This is a performance requirement of the optics. Low ensquared energy per sampling leads lower spectral resolution than expected, as well as increasing cross-talk between lights from adjacent fibers. Optics should be well designed to achieve required ensquared energy specification. Also, this requirement should be fulfilled over the entire field points since spectra of fibers are equally important.

- Average throughput > 35%

DOTIFS sensitivity requirement (emission line S/N ~ 5 [1]) from one-hour exposure sets throughput budget for each sub-system. Throughput higher than 35% for spectrograph optics is required to fulfill the budget. The throughput includes throughput of the collimator, filter, grating, camera and CCD quantum efficiency.

**2.2 Base design**

DOTIFS spectrograph optics is designed based on South African Large Telescope Robert Stobie Spectrograph (SALT RSS) optical design. We studied this design as one of starting design candidates and chose as final one because of its similar optical characteristic to requirements of DOTIFS spectrograph. Details of SALT RSS design including optics prescription are described in its optics design paper[4]. SALT RSS is an imaging spectrograph with long-slit and multi-slit spectroscopy and Fabry-Perot imaging spectroscopy capabilities. Unlike other conventional slit spectrographs, this spectrograph is designed for fast input beam (F/4.2). It also has a fast F/2.2 camera with wide a field of view as well as wide wavelength coverage. Since DOTIFS spectrograph also need to accommodate fast input beam (F/4) with fast camera optics (F/1.5), SALT RSS becomes the excellent starting point for designing DOTIFS spectrograph optics. Table 1 shows the difference between DOTIFS and SALT RSS parameters.

Table 1. DOTFIS vs SALT RSS optics parameters

| Parameter | DOTIFS | SALT RSS |
|---|---|---|
| Wavelength range | 370-740 nm | 320-900 nm |
| Collimator F/# | F/4 | F/4.2 |
| Camera F/# | F/1.5 | F/2.2 |
| Pupil diameter | 130 mm | 150 mm |
| Slit length | 80 mm | 107 mm |
| Camera field of view | 13.5° | 16° |
| Field size | 30 mm x 45 mm | 30 mm x 90 mm |

**2.3 Optics design procedure**

We use optical design software (Zemax) to design the spectrograph optics. In the beginning, we determine design requirements of the optics as listed in Table 1. After that, we set up merit function to satisfy given requirements and perform optimization. We optimize collimator and camera optics in order. During the optimization, we manually add or subtract lens surface or element. Detail design prescription of collimator and camera optics is given in chapter 3.

Table 2. DOTIFS subsystem main parameters

| Subsystem | Parameter | Value |
|---|---|---|
| **General** | Wavelength coverage | 370-740 nm |
| | Spectral resolution | 1200-2400, 1800 @555 nm |
| | Operation temperature | -5 to +22 °C |
| **Fiber Slit** | Fiber slit size | 80 mm |
| | Number of fibers per spectrograph | 288 |
| | Fiber model | CeramOptec WF100/110/125P12 |
| | Fiber core diameter at the slit | 100 μm |
| | Fiber output focal ratio | F/4.17 |
| **Collimator** | Focal length and focal ratio | 520 mm, F/4 |
| | Optics composition | 3 singlets and 2 close pair singlets |
| **Filter** | Filter range | 370 – 740 nm |
| | Tilt angle | 10° |
| **VPH Grating** | Line per frequency | 615 lines / mm |
| | Angle of incidence | 8.49° |
| **Camera** | Focal length and focal ratio | 195 mm, F/1.5 |
| | Optics composition | 3 singlets, 3 close pair singlets |
| **CCD** | Detector model | E2V CCD44-82 |
| | Detector format | 2048 x 4096 pixels |
| | Pixel size | 15 μm |

## 3. DOTIFS SPECTROGRAPH OPTICS SUBSYSTEM

In this chapter, we describe each subsystem of the DOTIFS spectrograph in detail. We also present design procedure of each system. Table 2 lists parameters of DOTIFS spectrograph subsystems.

### 3.1 Fiber Slit

In a fiber-fed spectrograph, fibers are located at the position of a slit in a conventional long-slit spectrograph and acting as a slit. This so-called fiber slit is composed of fibers with certain spacing in between (Figure 2). Light coming out from fibers is dispersed in the spectral direction, which is perpendicular to the slit direction. Therefore, spectrum from each fiber forms multiple line-like images on the CCD. Ideally, an image of fiber at specific wavelength should be looked like a finite circle, because fiber cores have a circular shape. However, the actual fiber images are bigger and fuzzier than the theoretical size due to aberration. It brings a phenomenon called fiber cross-talk, which is overlapping of light from two adjacent fibers on the CCD. Thus, separation between fibers should be wide enough to minimize this cross-talk. At the same time, the separation should be narrow as much as possible to accommodate the maximum number of fibers in a single spectrograph, to reduce the number of total spectrographs. Therefore, proper spacing between fibers is necessary to be investigated. It is known that this separation should be at least 2-3 times bigger than Full Width-Half Maximum (FWHM) of fiber image profile[5]. In DOTIFS spectrograph, fiber image size is corresponding to 2.5 pixels on CCD so the spacing should be between 5 to 7.5 pixels. We applied such spacing as 6.75 pixels, which is 270 μm separation at the fiber silt (Figure 2, left) so fibers can be properly separated within ~2000 pixels in the spatial direction of the CCD. This corresponds to 288 fibers per one spectrograph.

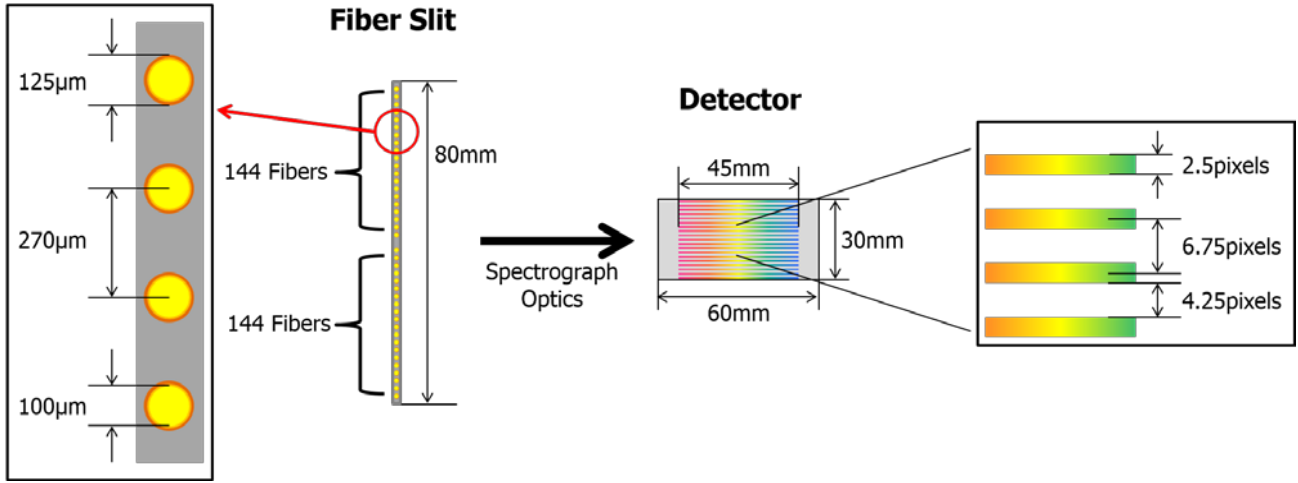

Figure 2 Conceptual diagram of fiber slit and detector. One fiber slit contains light from two IFUs. The arrangement of the fibers is shown at the enlarged diagram on the left. Light from each fiber is dispersed by spectrograph optics and forms spectrum on the CCD, as shown in the right side diagram. Each spectrum is separated by 6.75 pixels.

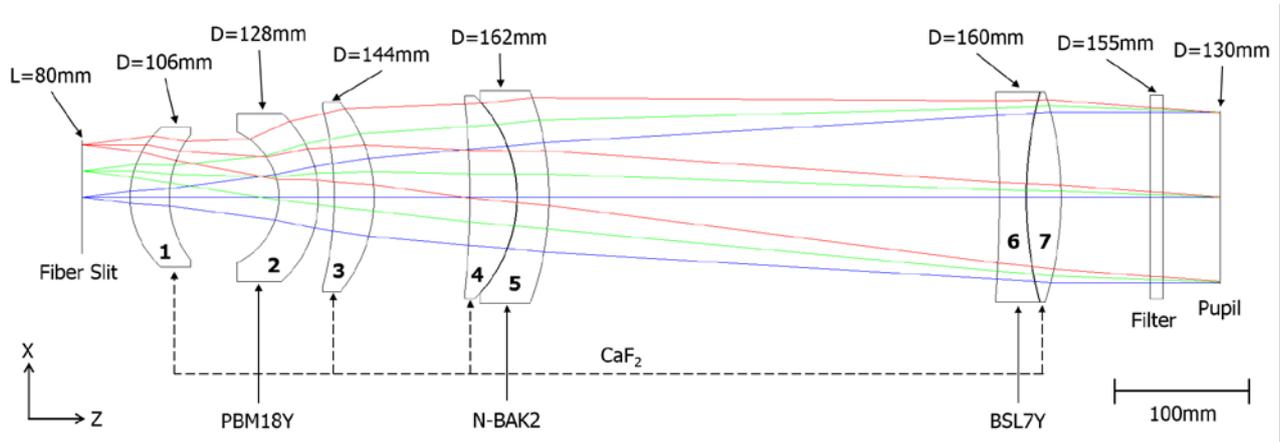

Figure 3 Collimator optics layout. Diameter and glass material of each component is shown. Different color lines represent light from different field points. Path of the light is shown for only three field points for clarification.

### 3.2 Collimator

We design DOTIFS collimator to refract F/4 light from 80 mm fiber slit into collimated 130 mm diameter beam in the entire working wavelength range. The requirement for near-UV wavelength range (370 nm-400 nm) makes this design challenge since only a few available glass materials have high transmission (better than 99.5 %/10 mm) in this range.

We take elements from the SALT RSS collimator, (F/4.2, 120 mm entrance slit, 150 mm diameter pupil, focal length=630 mm, 9 elements in 5 groups) as starting design and re-optimize to the DOTIFS requirements. Based on the measurement of the intensity distribution profile obtained from light coming out from actual optical fiber, we choose object property as NA=0.125 Gaussian profile beam with apodization factor of 3 for the collimator. A perfect 195 mm focal length paraxial lens is used during collimator optimization process. We set merit function as a sum of Root-Mean-Square (RMS) spot radius of 5 different wavelengths and 9 different field points. Also, the effective focal length is constrained to be 520 mm, and the incident angle of marginal rays at the pupil position is tried to be parallel as much as possible.

The design (Figure 3) starts with 3 singlets and followed by 2 pairs of closely separated singlets (7 elements in total.). We initially designed the close pairs as doublets, but later we decide to separate them to avoid possible risk during fabrication and delivery of optics. Compare to the starting design, the size and number of the lens had been decreased since our field and pupil diameter are smaller than the original. We use only spherical lenses to save fabrication cost as well as minimize alignment difficulty. Also, it has 2 pairs of closely separated singlets instead of 2 doublets + 1 triplet configuration in the starting design. Lens diameter is ranging from 106 mm to 162 mm.

### 3.3 Broadband Filter

We introduce the broadband filter to filter lights out of working wavelength range, to prevent second order contamination by wavelength shorter than 370nm and reduce stray light. Thus, we design the filter as a bandpass filter for the wavelength range from 370nm to 740nm. The filter is located next to the pupil and the dispersion element, so the light out of working wavelength cannot propagate beyond.

Also, we tilt the filter with respect to the slit direction (X-axis in Figure 3) by 10 degrees. This tilting is shown in Figure 4 but not in Figure 3, since Figure 4 is rotated by 90 degrees with respect to the optical axis compared to Figure 3. We tilt the filter to remove light reflected from the camera optics side. Without filter tilting, the light can be reflected again by the filter and produce stray light on CCD. Considering the path of light reflected by the different angle of the tilted filter, wavelength dependent efficiency of the grating, and maximum incident angle of the collimated beam to the filter, we determine the tilted angle as 10° to effectively minimize the stray light.

### 3.4 Pupil and Dispersion Element

Dispersion element is a heart of spectrograph which disperses lights of a different wavelength in different directions. It is located is 30 mm after the pupil location to minimize the incident of scattered light coming from the collimator side. Since DOTIFS working wavelength range and spectral resolution do not require exchangeable grating or dichroic, we adopt a single and fixed dispersion element in the design.

At the beginning of design phase, both conventional ruled grating and Volume Phase Holographic (VPH) grating had been investigated as a dispersion element. The ruled grating has an advantage on relatively wide bandwidth and flat efficiency curve, but it was turned out to be expensive and took long lead time for production. On the other hand, VPH grating has relatively higher throughput and can be designed with smaller optics since it is transmitted element[6]. Also, it is cheaper than ruled grating and can be manufactured within few months[7]. Based on the result of expected throughput comparison between both types of gratings, we choose VPH grating as a dispersion element of DOTIFS spectrograph.

Grating properties are determined by various parameters; dispersion, pupil diameter, camera F ratio and diffraction angle. For grating spectrograph configured at Littrow condition, wavelength dispersion on focal plane at the first order is given by

$$\lambda = 2 \times \sin(\theta_{Littrow}) / L, \quad \text{dispersion (nm/µm)} = d\lambda / dw = \cos(\theta_{Littrow}) / L / f_{camera}$$

where $\theta_{Littrow}$ is angle of the Littrow condition, w is length in the spectral direction, L is line density of a grating, and $f_{camera}$ is focal length of a camera optics. Since dispersion value is set as 0.00822 nm/µm from spectral resolution requirement, other parameters should be determined appropriately to yield the dispersion.

We consider a few things while determining those parameters. First, choice of angle of incidence should be similar to the Littrow wavelength, which is a wavelength with the highest efficiency over entire wavelength coverage. Also, a high angle of incidence to the grating increases the dispersion as well as the size of the grating. Second, the focal length of camera depends only on pupil diameter because F ratio is already fixed as F/1.5. In practice, design optics with larger pupil size is easier than smaller since larger pupil size leads longer focal length for both collimator and the camera with given F-number. However, larger pupil also makes optics bigger so that it increases weight and fabrication cost. Third, lines per mm in VPH should be as low as possible to achieve broad bandwidth on response curve since our wavelength coverage is relatively wide for VPH grating[8]. At the same time, it is also required to be dense enough for manufacturability. This gives about 600 lines/mm as a minimum grating line density.

By considering above characteristics of parameters, we decided the final parameters of the grating and size of the pupil. We use Rigorous Coupled-Wave Analysis (RCWA) simulation software, GSolver (Version 5.2) to explore optimal grating parameters which yield reasonable optics size and high grating efficiency. The final parameter is determined as in Table 2. We choose Littrow wavelength as 480 nm instead of the central wavelength (555 nm). This is due to known phenomena called Littrow ghost[9], which is appeared as a bright spot on the image plane at the location of Littrow wavelength. Since there are many known galaxy emission/absorption lines just before 555 nm, there is a chance that one of those emission lines from redshifted galaxy can be overlapped with the ghost and cause a problem, if we do not remove it. Although there is some known way to remove this ghost[9], the methods are not easily applicable in case of DOTIFS spectrograph. Therefore, we decided to shift the ghost wavelength to mitigate the problem rather than removing it. By shifting Littrow wavelength to 480 nm, which is shorter than known emission/absorption lines, we can minimize the overlapping of the ghost with those known lines. In this case, lines longer than 480 nm will never be suffered by the ghost, and lines shorter than 480 nm is located quite far from there, (e.g., OIII: 436.32 nm, H-gamma: 434.1 nm) thus they will also not be contaminated by nearby galaxies with redshift less than z=0.1, which are primary target of DOTIFS.

### 3.5 Camera Design

DOTIFS camera optics is designed to focus the dispersed light on the CCD. The camera is articulated about the grating axis to choose the wavelength range to focus. We choose the angle between grating axis and camera axis as 11.17 degrees so that the central wavelength light will pass through the optical axis of the camera. The F ratio of the camera is determined as F/1.5 (focal length = 195 mm) by the collimator F-number, fiber core size, pixel size and the spectral resolution requirement. This fast F-ratio makes this design challenging and sensitive to tolerances. Optics size should be large enough to accommodate whole light coming from the dispersion element within working wavelength range without any vignetting.

Again, starting design takes elements from the SALT RSS camera (F/2.2, 95 x 61 mm focal plane, focal length=330 mm, 9 elements in 4 groups), and re-optimized to fulfill the DOTIFS requirements. For initial optimization, a perfect paraxial lens (focal length = 520 mm) is used as an artificial collimator. After the initial optimization, the paraxial lens is replaced with the actual collimator optics design which is designed as in the Ch. 3.2. The merit function minimized the RMS spot radius in the same way as of the collimator. An effective focal length is constrained to be 195 mm.

The design (Figure 4) starts with 3 pairs of closely separated singlets and followed by 3 singlets (9 elements in total.) Compare to the starting design, size of lens elements are decreased, but some elements remain identical although the size of the focal plane is smaller than the original. This is because the starting design has an aspheric surface, but our design has only spherical surfaces. Also, F-ratio of our design is faster than the original. Thus it demands more surfaces to achieve desired performance. Field of view of the camera is about 13.5°.

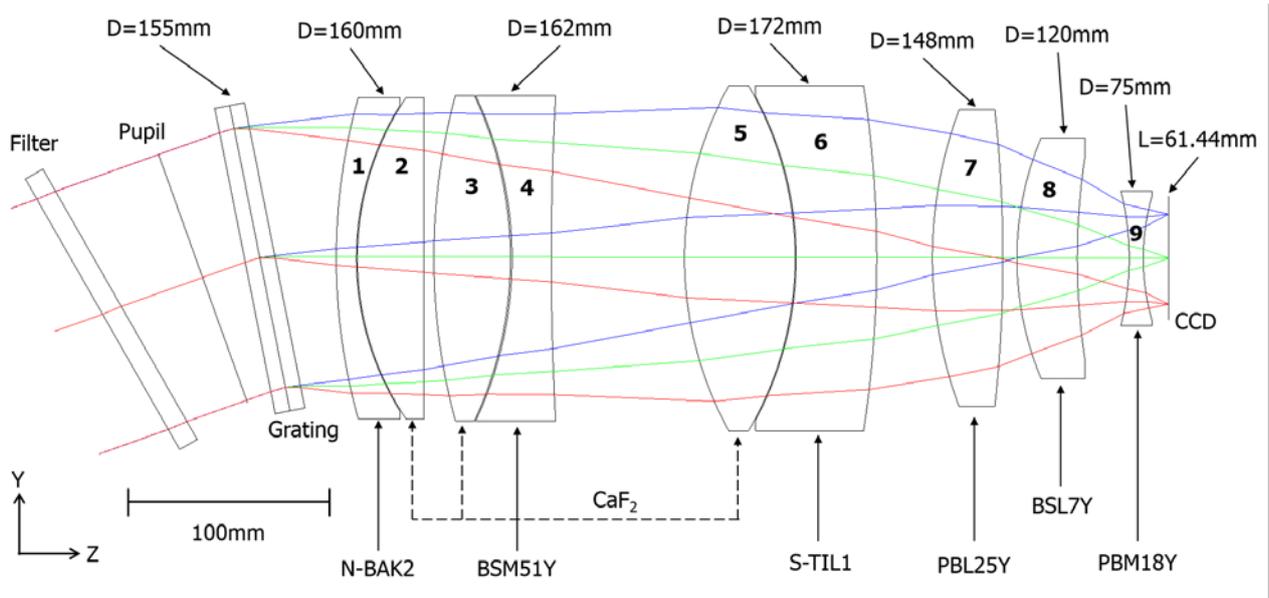

Figure 4 Camera optics layout. Descriptions are identical as in Figure 3. In this figure, the color of light represents difference wavelength. Blue is 370 nm, and red is 740 nm light. A tilted filter with respect to the optical axis is shown in this figure.

### 3.6 Detector

The detector size is chosen based on spectral resolution requirement, CCD availability, and the capability of the optics. The size in spectral direction is set as 3,000 pixels to accommodate the whole wavelength coverage with required spectral resolution (See 2.1). Since there is no available CCD with 3K in one direction from major CCD manufacturers, we have to choose 4K in spectral direction and remains 1K as unused. The size in the spatial direction is determined to minimize expected cost with reasonable complexity. It gives us two options on the size of the spatial direction, 2K pixels or 4K pixels. 4K by 4K CCD requires bigger and more optical elements, but it can reduce the number of the spectrograph. On the other hand, 2K by 4K requires smaller optics, but double the number of the spectrograph to accommodate the same number of total fiber compare to 4K by 4K CCD option. After estimation of fabrication cost and weight budget of both options, 2K by 4K CCD option with 8 spectrographs is turned out to be the more economical solution. The final decision on CCD model is e2v CCD44-82 chip (2048x4096, 15 μm size pixels).

We adopt new coating technique on CCD to increase throughput. DOTIFS spectrograph has fixed wavelength distribution on the CCD since the spectrograph is in a single configuration with fixed dispersion element and working wavelength range. We take this as an advantage to maximize the throughput. Unlike the traditional broadband coating, we applied wavelength dependent coating called graded coating along the spectral direction, so coating on each pixel will be optimized for the wavelength of light which will fall on the pixel. Wavelength map of our coating scheme is shown in Figure 5. This graded coating improves CCD quantum efficiency by several percentage points, especially in the blue and near UV region.

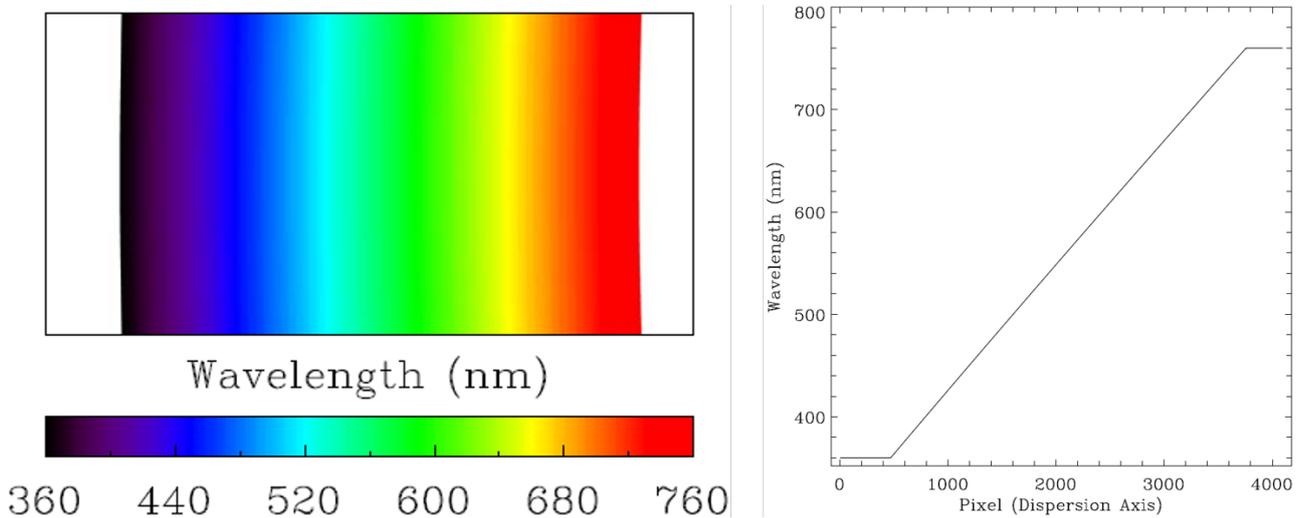

Figure 5 (Left) Wavelength map of the graded coating on the CCD proposed to the vendor. Outer black boundary represents CCD detector. Left and right sides are left as white since those are the unused area. (Right) Optimized wavelength depends on CCD pixel in dispersion axis.

## 4. TOLERANCE AND THERMAL ANALYSIS

### 4.1 Tolerance Analysis

We performed tolerance analysis with Zemax optical design software. This is to find out required fabrication and alignment tolerance of DOTIFS spectrograph optics. Requirements for lens fabrication and alignment accuracy are derived and stated in Table 3 and Table 4. We did sensitivity and Monte-Carlo analysis with those requirements as well. Results show that we can achieve the tolerance requirements by modern fabrication and alignment technology. Since DOTIFS spectrograph uses the distance between the last and the second last element of the camera optics as a compensator, we use the same for this analysis.

We present the result of sensitivity analysis in Table 5. The change of average RMS spot radius is 5.64 μm. In general, camera optics is more sensitive than the collimator due to its fast F-ratio. Major worst offenders in fabrication requirements are the surface tilt of close pair singlets in camera optics. The performance is also sensitive to refraction index and Abbe number of several camera elements. It shows similar trends in the sensitivity analysis of alignment requirements. Element tilt of close pair singlets in camera is the most sensitive offender among alignment tolerances.

We ran 100,000 Monte-Carlo simulations, assuming normal distribution statistics with requirements above. Results are stated in Table 6. The statistics show that mean RMS spot radius as 11.1 μm, with a standard deviation of 3.1 μm. This difference is considered to be acceptable. Additional analysis on designs generated by Monte-Carlo simulation also confirms that the result is satisfying optics performance requirements.

Table 3. Optics lens fabrication requirements

| Center thickness | ±50 μm |
|---|---|
| Radius of curvature | ±0.1% |
| Surface tilt | ±2 arcmin |
| Surface decenter | ±50 μm |
| Surface irregularity | λ/4 @ 632.8 nm |
| Refraction index | ±0.0003 |
| Abbe No. | ±0.5% |

Table 4. Optics alignment accuracy requirements

| Center distance | ±50 μm |
|---|---|
| Element tilt | ±2 arcmin |
| Element decenter | ±50 μm |

Table 5. Sensitivity analysis result

|  | Design | Fabrication | Alignment | Total |
|---|---|---|---|---|
| RMS spot radius (μm) | 5.79 | 10.44 | 8.99 | 11.43 |
| Change (μm) |  | 4.65 | 3.20 | 5.64 |

Table 6. Monte-Carlo analysis result

| RMS spot radius (μm) |  | Compensator (mm) |  |
|---|---|---|---|
| Nominal | 5.8 | Minimum | 25.261 |
| Worst case | 35.4 | Maximum | 26.293 |
| Mean | 11.0 | Mean | 25.807 |
| Standard deviation | 3.0 | Standard deviation | 0.128 |

**4.2 Thermal Analysis**

We performed thermal analysis to estimate the impact of temperature variance on optics and compensate performance degradation derived from the variance. Since the spectrograph optics will be thermally equivalent with the surrounding environment, the effect of the temperature variance should be clarified. The spectrograph optics is designed and optimized for the temperature of 20°C. Air temperature at the observing site is known as varying between -5 to +22°C yearly while the intra-night temperature variation is less than 2°C[2]. Although thermal effect within a single night is expected to be negligible, the variance over entire year should be considered because of its wide temperature variation range and $CaF_2$ elements which has a large value of temperature dependent index of refraction variance and thermal expansion coefficient. In general, performance degradation due to temperature variance can be easily compensated by changing the distance between two elements in the optics. As in the tolerance analysis, we use the distance between the last and the second last element of the camera optics as a compensator.

We again ran thermal analysis with Zemax optical design software. In Zemax, size, thickness, and index of refraction of optics are changed by temperature variation. We use thermal index variation and expansion coefficient data of each glass from catalogs supplied by glass manufacturers. The thermal coefficient of Aluminum 6061-T6, (23.5 x $10^{-6}$/°C) is used for expansion of opto-mechanical structure in between. We consider only expansion along the optical axis in this analysis. We performed thermal analysis in two conditions. First, we tested the effect of temperature change among annual temperature variation range at the telescope site. This analysis had been done with temperature range from -5 to +25°C with 5°C increments. At each temperature, first we measure RMS spot radius without any compensation. Then we tried to re-optimize optics to minimize RMS spot radius using compensator and record corrected RMS spot radius as well as the displacement of the compensator. The result shows that the temperature variation significantly changes the RMS spot radius. At the same time, it also shows a compensator with a reasonable displacement can successfully correct such change (Figure 6, Left). Second, we investigated the impact of temperature variance within intra-night variance range. We select two nominal temperatures. (0 and 15°C) At each temperature, we first re-optimize optics using compensator, and measure its RMS spot radius changes over +/-1.5°C with 0.5°C increments without any additional compensation. This procedure assumes an actual observing condition, which will do optics compensation only at the beginning of each night. Result shows that RMS spot radius variation over intra-night temperature variance at both nominal temperature is very small. (Figure 6, Right) Thus, once the optics is compensated at the beginning of each night, there will be no significant performance degradation during rest of the night. Compensator mechanics is introduced in the opto-mechanical structure by putting camera dewar on the high precision linear stage. Image quality optimization algorithm will be developed and included in the instrument control software so that this procedure can be done automatically. The algorithm some continuum images at different stage location, then find out optimal displacement which has minimum line width along the spatial direction.

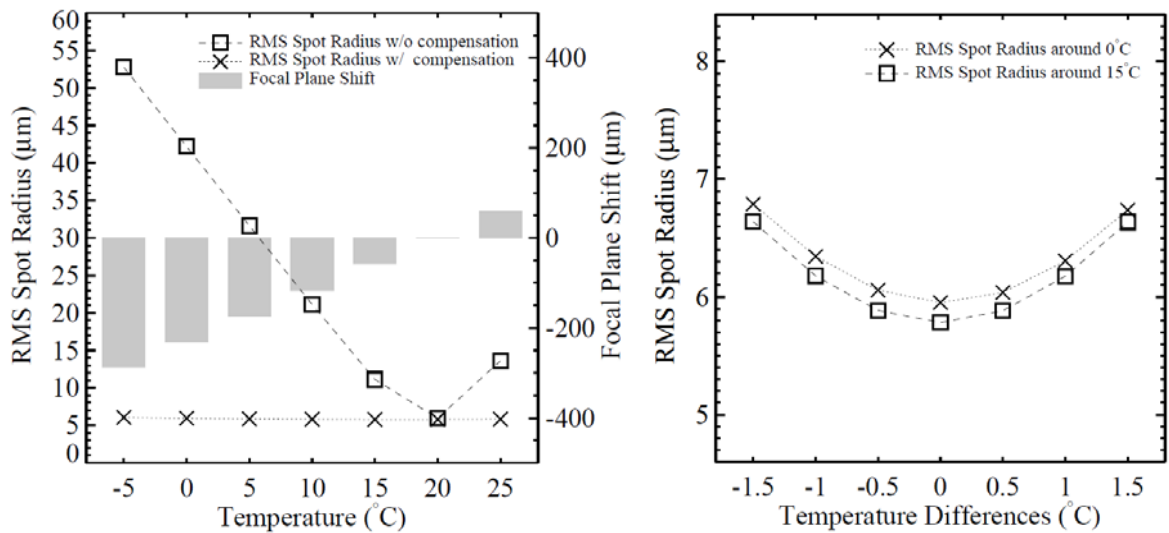

Figure 6 (Left) Result of thermal analysis with and without compensation within a yearly temperature range of the telescope site. (Right) Similar result without compensation within the intra-night temperature range.

## 5. DOTIFS SPECTROGRAPH OPTO-MECHANICAL DESIGN

We design opto-mechanical structure of the spectrograph to hold spectrograph optics rigidly independent of orientation with respect to the gravity. The result of spectrograph optics tolerance analysis is applied on the opto-mechanics design so that the design should fulfill its tolerance requirements. Circular holders are used to supporting collimator and camera enclosures. Individual lens holders are located inside of the enclosures to hold the lenses at their exact location properly. We use the linear stage to move CCD unit, and use the distance between the last camera element and the second last element as a compensator. Figure 7 shows the overall structure of the spectrograph opto-mechanics. The spectrograph is also designed to be attached to instrument frame at the Cassegrain direct port. In total 8 spectrographs occupy space around Cassegrain direct port instrument, ADFOSC (Figure 1). There will be two spectrographs at each side, in total 8 spectrographs along 4 sides. The entire structure is designed to mount spectrograph easily.

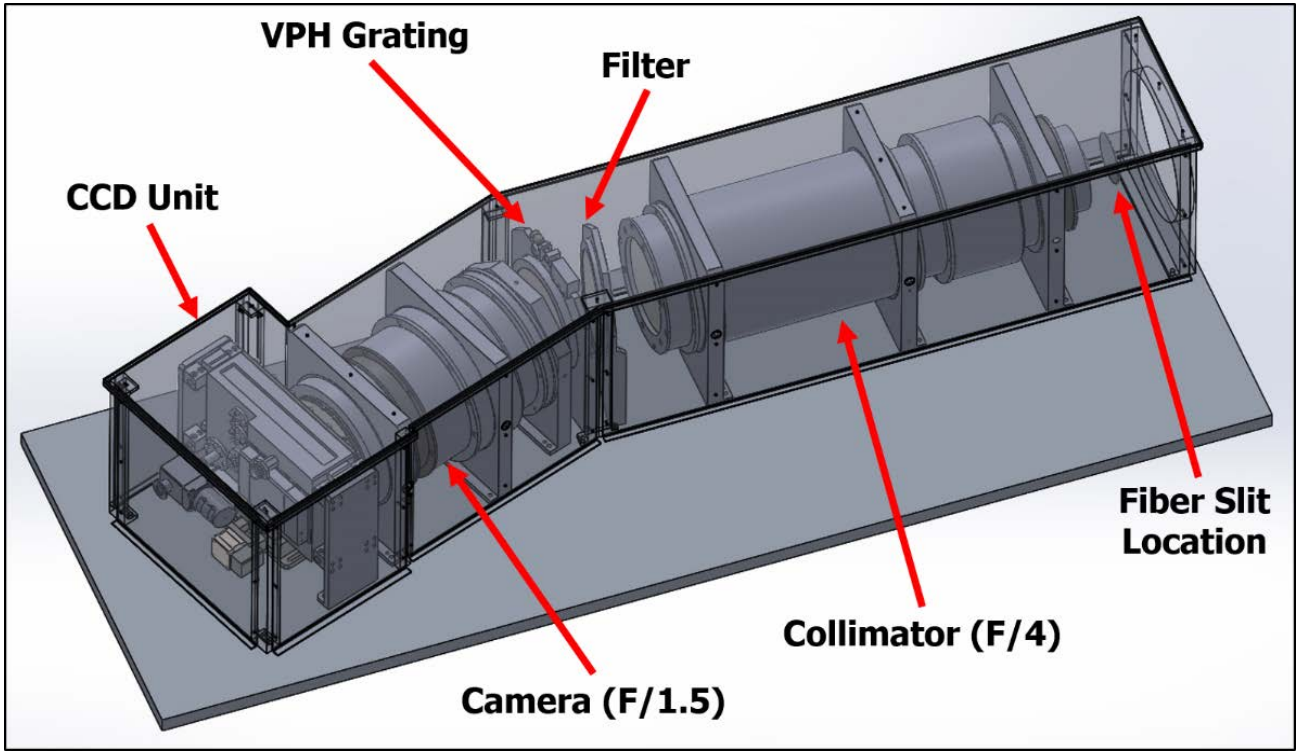

Figure 7 Opto-mechanical structure of DOTIFS spectrograph. Location of sub-components is pointed out with red arrows. CCD dewar is not shown in this figure.

## 6. PERFORMANCE

In this chapter, we present optical performance of the spectrograph optics in several aspects. We also developed a data simulator which simulates CCD image of the spectrograph. Details regarding the simulator can be found in a separate paper[10].

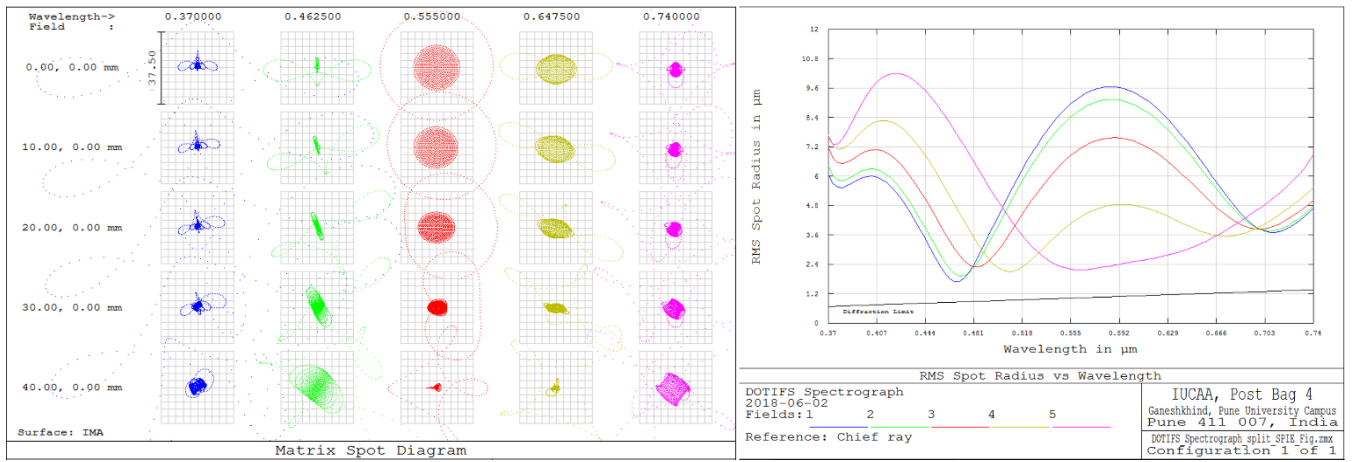

Figure 8 (Left) Matrix spot diagram of the spectrograph optics. Color represents different wavelength. (Right) Wavelength versus RMS spot radius plot. Color represents the light of different field points, as in the left figure from top to bottom rows.

## 6.1 Spot Diagram

In Figure 8, we present spot diagram and wavelength versus RMS spot radius graph of the spectrograph optics. The box size of the spot diagram is 37.5 μm which correspond to 2.5 pixels on CCD. Although the diagram shows quite a large spot shape at their out-most region, most of the lights are concentrated at the center. Therefore it has very small RMS spot radius as appeared in the wavelength versus RMS spot radius graph.

## 6.2 Extended Source Ensquared Energy

Extended source ensquared energy is measured to check fulfillment of the design goal. The design goal was higher than 80%, and the result satisfies the goal. (Figure 9)

## 6.3 Spectrograph Throughput

We present throughput of the spectrograph optics in Figure 10. Throughputs of the individual sub-component are plotted with different colors. Due to high-throughput optical glasses, throughput loss from collimator and camera optics are mainly attributed from AR coating, not from glass except near-UV range. Throughput curve of CCD response is from the theoretical expectation of CCD quantum efficiency with the graded coating. Similarly, the curve of VPH grating is also from the theoretical simulation, fitted to the several measured values. The curve of the broadband filter is again from the theoretical simulation. Thus, the actual curves of CCD, VPH grating and filter could be different from what shown here. The black line shows combined throughput of spectrograph optics. It is 53.9% at peak wavelength and 41.8% on average, which satisfies the design goal of the optics (> 35%). We also show throughput of the entire DOTIFS instrument which includes fore-optics, IFU, fibers, and spectrograph, but without sky transmission. Average throughput of the instrument is 26.0%.

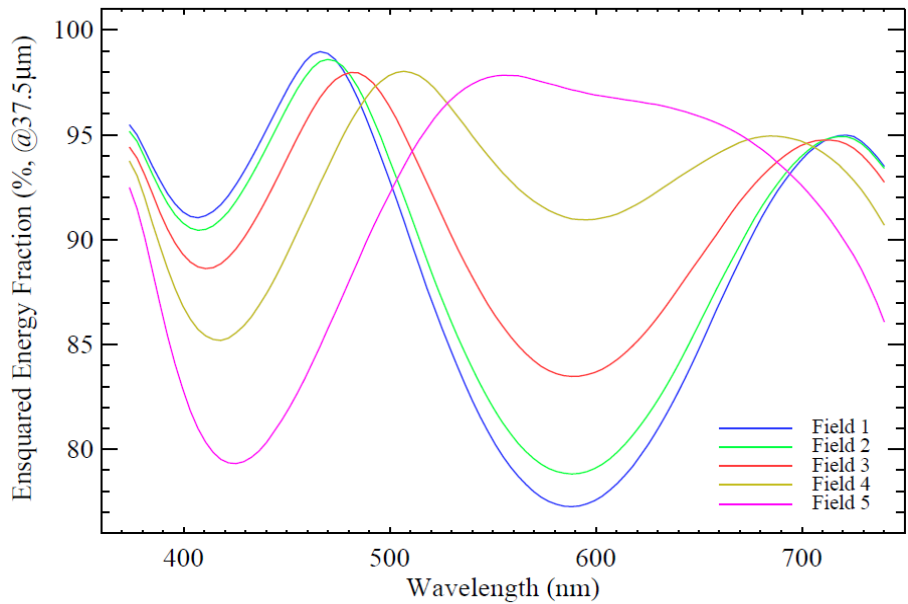

Figure 9 Extended source ensquared energy fraction of the spectrograph optics with 100 μm diameter, NA=0.125 uniform and circular light source. Sampling size is 37.5 μm, which corresponds to 2.5 pixels size. Color represents energy fraction from different field points, as in the right side of Figure 8.

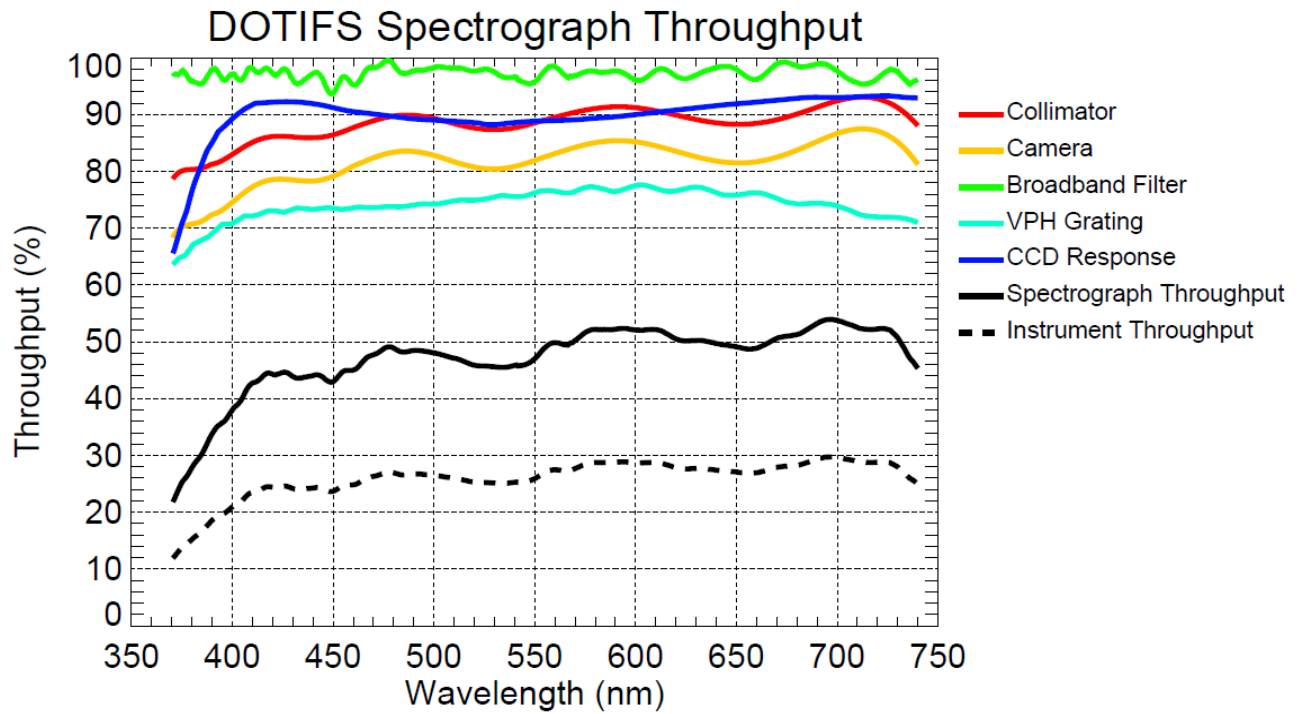

Figure 10 DOTIFS spectrograph throughput

## 7. CURRENT STATUS

We are planning to assemble two spectrographs first and commission the instrument, and add six more spectrographs in the future to complete the instrument. As of 2018 May, we finalized spectrograph optics design and ordered two sets of lens components to the lens fabricator (Phoenix Optical Technologies, UK). We expect to receive actual elements within 2018. We finalize the opto-mechanical structure design and complete the parts fabrication at the local machine shops except for some parts for baffling. The parts will be delivered to the laboratory before the arrival of the optics. We ordered broadband filter (Asahi Spectra, Japan) and VPH grating (Wasatch Photonics, USA) for two sets of the spectrograph and received them before. Currently, we are doing performance verification of received components. Order of CCDs are also completed, and we receive them in 2017. We will start assembly of spectrograph optics when all parts are arriving. We are working on other parts of the instrument as well. Optics for instrument fore-optics were ordered along with the spectrograph optics. We made proto-type fiber array and tested characteristic of microlens array which will be part of IFU. Actuators for IFU deployment system are tested underway.

## 8. SUMMARY

We present the optical and opto-mechanical design of the DOTIFS spectrograph. F/4 collimator and F/1.5 camera optics are designed based on the prescription of SALT RSS spectrograph. Design procedures and behind rational of the spectrograph are described. Expected performance of the spectrograph fulfills the design requirement very well. Currently, we ordered and received most of the parts except lens elements. We will start spectrograph assembly once we have all parts on our hands.